\documentclass[11pt, letterpaper]{article}

\usepackage{graphicx}
\usepackage{amsmath}
\usepackage{amssymb}
\usepackage{mathtools}
\usepackage{setspace}
\usepackage{enumitem} 
\usepackage{epstopdf}
\usepackage{color}
\usepackage[letterpaper,left=1in,right=1in,top=1in,bottom=1in]{geometry}
\usepackage[]{algorithm2e}
\usepackage{multirow}
\usepackage{longtable}
\usepackage{rotating}
\usepackage{mathrsfs}
\usepackage{subfig}
\usepackage{float}
\usepackage{graphicx}
\usepackage{caption}

\graphicspath{{Figures_new}}

\title{\Large \bf Soil moisture map construction using microwave remote sensors and sequential data assimilation}

\author{
\centerline{\normalsize Bernard T. Agyeman$^{a}$, Song Bo$^{a}$, Soumya R. Sahoo$^{a}$, Xunyuan Yin$^{a}$, Jinfeng Liu$^{a,}$\thanks{Corresponding author: J. Liu. Tel: +1-780-492-1317. Fax: +1-780-492-2881. Email: jinfeng@ualberta.ca.}, Sirish L. Shah$^{a}$}
\vspace{5mm}\\
\centerline{\small $^{a}$Department of Chemical \& Materials Engineering, University of Alberta,}\\
\centerline{\small Edmonton, AB T6G 1H9, Canada.}}
\allowdisplaybreaks

\begin{document}

\date{}
\maketitle

\setstretch{1.5}

\begin{abstract}
Microwave remote sensors mounted on center pivot irrigation systems provide a feasible approach to obtain soil moisture information, in the form of water content maps, for the implementation of closed-loop irrigation. Major challenges such as significant time delays in the soil moisture measurements, the inability of the sensors to provide soil moisture information in instances where the center pivot is stationary, and the inability of the sensors to provide soil moisture information in the root zone reduce the usability of the water content maps in the effective implementation of closed-loop irrigation. In this paper, we seek to address the aforementioned challenges and consequently describe a water content map construction procedure that is suitable for the implementation of closed-loop irrigation. Firstly, we propose the cylindrical coordinates version of the Richards equation (field model) which naturally models fields equipped with a center pivot irrigation system. Secondly, measurements obtained from the microwave sensors are assimilated into the field model using the extended Kalman filter to form an information fusion system, which will provide frequent soil moisture estimates and predictions in the form of moisture content maps. The utility of the proposed information fusion system is first investigated with simulated microwave sensor measurements. The information fusion system is then applied to a real large-scale agriculture field where we demonstrate the its ability to address the challenges. Three performance evaluation criteria are used to validate the soil moisture estimates and predictions provided by the proposed information fusion system.

\end{abstract}


\section{Introduction}

Over the last century, the world has witnessed a rapid increase in its population: from 1.8 billion in 1915 to 7.5 billion in 2017~\cite{srivastava2018}. The world's population is projected to increase by 31\% from its current estimate of 7.7 billion to 9.8 billion by 2050~\cite{pRB2016}. Over the same period, average per capita income is also expected to rise~\cite{Wallace2000}. Jointly, these two developments are driving up the global demand for food. 
Freshwater and arable land constitute two of the most fundamental resources for food production. There is however a restriction on cropland expansion in the wake of the high global demand for food due to the high ecological and social trade-offs of clearing more land for agriculture~\cite{ART2016}. With a limited potential to increase suitable cropland, agricultural irrigation has become an increasingly important tool to ensure sufficient global supply of food in the near future~\cite{Wichelns2006}. The expansion of irrigation will put further stress on rivers, lakes, and aquifers that are already depleted, raising concerns about the ability of the Earth to feed humans with its limited freshwater resources. According to the United Nations, agriculture accounts for more than 70\% of anthropogenic water withdrawals, with the main consumer being irrigation~\cite{UN_report}. Currently, the water-use efficiency in irrigation is about 60\%~\cite{WWDRUN2009}, which means that a significant portion of the water used in irrigation is wasted due to inefficient irrigation practices. Ensuring the near-optimal use of water through increased water-use efficiency in irrigation represents a popular and effective approach to reduce water losses in agricultural irrigation, thus mitigating the current water supply crisis.

Currently, most irrigation systems are implemented in an open-loop fashion. In an open-loop system, the operator makes the decision on the amount of water to be applied and the timing of the irrigation event. These decisions are made based on the operator's empirical knowledge of the crop's water requirements, soil type, and climatic factors. This one-way information flow approach often leads to excessive consumption and wastage of water resources. Closed-loop irrigation is a promising alternative to improve the water-use efficiency in irrigation and consequently reduce the consumption and wastage of water resources. In closed-loop systems, the operator develops a control strategy that takes over and makes detailed decisions on when to apply water and how much water to apply. Irrigation decisions are made and actions are carried out based on real-time soil moisture information as well as climatic and crop information. In this type of system, feedback and control of the system are done continuously.

To implement closed-loop irrigation, various moisture sensing techniques must be employed to provide the required soil moisture information for feedback control. In-situ moisture probes (point sensors) are the most mature and reliable sensing techniques for soil moisture. These sensors are portable, easy to install and operate, and can provide continuous measurements at various depths of the field. However, it is very expensive to obtain a thorough water distribution of a large-scale field using these probes~\cite{petropoulos2015surface,gupta2016integrative,shi2015parameter}. A significant number of closed-loop irrigation systems have been implemented based on real-time soil moisture information from point sensors. In~\cite{saavoss2016yield}, soil moisture measurements from a network of EC5 point sensors were used in the implementation of a closed-loop irrigation system for a greenhouse. A zone model predictive controller, and a closed-loop  controller and a scheduler were developed using point sensor measurements in~\cite{smao2018soil} and~\cite{nahar2019closed} respectively.
Non-invasive sensing techniques provide a more practical approach to capture the spatial variability of soil moisture. Some techniques in this category include ground penetrating radars, electrical resistivity tomography, electromagnetic induction, and microwave remote sensing. Microwave remote sensors have been demonstrated to have the capability of measuring the near-surface soil moisture quantitatively, providing soil moisture information for feedback control. Microwave sensors mounted on center pivots provide a feasible approach to obtain the soil water information of fields equipped with center pivots irrigation systems. Center pivot irrigation is a mechanized irrigation system that irrigates crops in a circular pattern around a central pivot. It is made up of a radial pipe supported by towers that pivot around a center point. It consists of equally spaced nozzles along its radial pipe and as it rotates, water is released from the nozzles and irrigates crops. As the center pivot rotates, microwave radiometers measure the soil moisture content of yet to be irrigated locations of the field, typically within a distance of between 10 to 20 m from the presently irrigated location and provide the soil moisture information in the form of water content maps at the end of the pivot's rotation cycle. In this approach, raw measurements obtained from the radiometers are extrapolated and interpolated to obtain unknown moisture content information at unmeasured portions of the field such as areas beyond the circular track of the pivot and areas between the tracks of the radiometers.
There however exist three main challenges associated with this sensing approach which reduce the utility of the water content maps in the implementation of closed-loop irrigation. Firstly, depending on the size of the field, it takes between two to three days for a center pivot to complete one rotation cycle. Hence, it also takes between two to three days to obtain a single water content map from the microwave sensors. Accordingly, the moisture content maps contain soil moisture information obtained from separate days within the rotation of the center pivot, an occurrence which results in time delays in the moisture measurements. The presence of time delays in the measurements means that continuous feedback and control cannot be effectively implemented. Secondly, the current approach lacks a predictive capability in the sense that the radiometers cannot provide soil moisture information of the field when the center pivot is stationary. Thus, before the start of the center pivot's irrigation cycle, this approach is unable to provide soil moisture information for feedback control. Lastly, this approach provides near-surface measurements of soil moisture which means that it fails to provide soil moisture information in the root zone, knowledge of which is required for the implementation of feedback control.

Observational data provided by the various soil moisture sensing techniques alone are not sufficient to provide accurate soil moisture information needed for feedback control since they are limited in their spatial and temporal coverage of soil moisture. Agro-hydrological models, such as the Richards equation, on the other hand, can provide spatially and temporally continuous predictions of soil moisture at specified space and time resolutions, and at various soil depths. The Richards equation is a partial differential equation (PDE) that describes the flow of water through unsaturated porous media under the action of gravity and capillarity. However, due to limited process knowledge and simplifications, numerical models are unable to predict soil moisture accurately. By integrating agro-hydrological models with observational data, observational gaps can be filled and improved predictions of soil moisture can be obtained. The process of integrating observed data with numerical models is known as sequential data assimilation. The primary components of a sequential data assimilation system include observations, numerical modeling, and an analysis update. The observations used for large-scale soil moisture data assimilation systems include observations from point soil moisture sensing techniques or remote-sensing observations. The numerical modeling consists of two parts: (i) a predictive field model, such as the Richards equation, to dynamically propagate the state and (ii) an observation model, such as the closed-form volumetric moisture content - capillary pressure head relationship provided by the van Genuchten model, to translate the state into the observed quantities. The analysis update combines the information from the observations and the model in an optimal fashion. Popular methods for the sequential assimilation of soil moisture observations include the extended Kalman filter (EKF)~\cite{lu2011dual,sabater2007near}, the ensemble Kalman filter (EnKF)~\cite{reichle2002hydrologic, Medina2014,moradkhani2005dual,chen2015comparison}, the particle filter~\cite{montzka2011hydraulic,pan2008estimation} and the moving horizon estimator (MHE)~\cite{bo2020parameter,bo2020decentralized}. In~\cite{reichle2002hydrologic}, soil moisture estimates are obtained by assimilating L-band (1.4GHz) microwave radiobrightness observations into a land model using the EnKF. In~\cite{bo2020parameter}, simultaneous soil moisture and parameter estimation was based on assimilating tensiometer measurements into the 1D Richards equation using the MHE. Bo and Liu~\cite{bo2020decentralized} proposed a decentralized framework for simultaneous soil moisture and parameter estimation based on the assimilation of capillary pressure head observations into the 3D Richards equations using the MHE.

This paper seeks to develop a moisture map construction procedure that is suitable for the implementation of closed-loop irrigation by addressing the aforementioned challenges. To eliminate the delays in the measurements, we propose the construction of frequently updated moisture content maps, with an update frequency up to a few minutes. Thus, instead of waiting for between two to three days before constructing a single water content map, a moisture content map will be constructed any time measurements are obtained from the microwave sensors. As new measurements become available, the previous map will be updated. A predictive capability will be incorporated into the current approach by proposing a suitable agro-hydrological framework, also known as the field model, which can make soil moisture predictions based on the prevailing field conditions, in instances where the center pivot is stationary. Finally, by integrating the microwave sensor measurements with the field model, soil moisture information at greater depths below the field surface, and hence the root zone can be estimated. To address these challenges, this paper seeks to meet the following specific objectives: 1) to develop an agro-hydrological framework (field model), that naturally models fields equipped with center pivots; 2) to assimilate microwave sensor measurements into the field model using a state estimation method to form an information fusion system; and 3) to construct moisture content maps using soil moisture estimates and predictions provided by the information fusion system. The main contributions of this paper include:

\begin{enumerate}
\item The cylindrical coordinates version of the Richards equation and a numerical scheme to solve the resulting equation. The cylindrical coordinates version of the Richards equation naturally describes an agro-hydrological system that is equipped with a center pivot irrigation system. 
\item  The assimilation of remotely sensed soil moisture obtained from microwave radiometers into the cylindrical coordinates version of the Richards equation using the extended Kalman filter to form an information fusion system.
\item The application of the proposed information fusion system to a real large-scale agriculture field. In this application, we demonstrate the utility of the information fusion system in dealing with the earlier outlined challenges. An approach to evaluate the performance (accuracy and consistency of the soil moisture estimates and predictions) is also discussed. The results of the performance evaluation indicate that the proposed information fusion system provides accurate and consistent soil moisture predictions when applied to a real case study.  
\end{enumerate}
\vspace{-2mm}
Some preliminary results of this paper were reported in~\cite{agyeman2021soilmoisture}. Compared with~\cite{agyeman2021soilmoisture}, this paper presents significantly detailed explanations, extended simulation results for different field conditions, and extensive performance evaluation of the proposed information fusion system.

\section{Model Development}
\subsection{Agro-hydrological System}
\begin{figure}[H]
	\centering
	\centerline{\includegraphics[trim={0cm 0cm 0cm 0cm},width=0.5\textwidth]{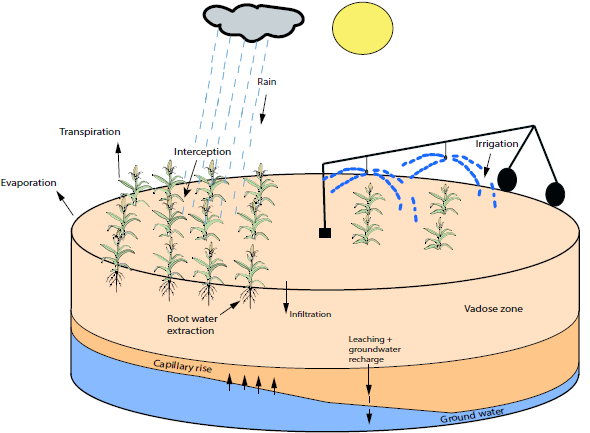}}
	\caption{An agro-hydrological system.}
	\label{fig:Polar_Agrohydrological}
\end{figure}

In this paper, we consider an agro-hydrological system that details the movement of water between crops, the soil, and the atmosphere. Figure \ref{fig:Polar_Agrohydrological} provides a simple illustration of an agro-hydrological system.

For irrigation purposes, the region above the water table, known as the vadose zone, is usually considered. In an agro-hydrological system, water transport takes place primarily through irrigation, rain, evaporation, transpiration, infiltration, root water extraction, surface run-off, and drainage. The main driving forces for water transport in soil are the capillary and gravitational forces. The transport  of water in soil can be modeled using the Richards equation \cite{richards1931capillary}. The Richards equation can be expressed in a general form as:

\begin{equation}\label{eq:Richardseqn_general}
\frac{\partial \theta}{\partial t}=C(h)\frac{\partial h}{\partial t}=\nabla.~(K(h)\nabla~(h+z)) -S(h,z)
\end{equation}
In Equation (\ref{eq:Richardseqn_general}), $h ~(\text{m})$ is the pressure head, $\theta ~(\text{m}^3 \text{m}^{-3})$ is the volumetric water or moisture content, $t~(\text{seconds})$ represents time, $z~(\text{m})$ is the spatial coordinate, $K(h)~(\text{ms}^{-1})$ is the unsaturated hydraulic water conductivity, $ C(h)~(\text{m}^{-1})$ is the capillary capacity, and $S\left(h,z\right)~(\text{m}^3 \text{m}^{-3}\text{s}^{-1})$ denotes the sink term.

\subsubsection{Root Water Extraction Rate}
Physically, the sink term, $S\left(h,z\right)$ in Equation (\ref{eq:Richardseqn_general}) represents the root water extraction rate which is defined as the volume of water taken up by the roots of crops per unit bulk volume of soil per unit time.
To solve the Richards equation, a function representing $S\left(h,z\right)$ is required. Simple empirical models have been developed to describe the root water extraction rate. In this paper, the Feddes model is used to represent $S\left(h,z\right)$. In this model, the root water extraction rate is computed at the various depths of the soil using the potential transpiration rate, the rooting depth, and the local prevailing soil water pressure head. Mathematically, the Feddes model~\cite{feddes1982simulation} defines $S\left(h,z\right)$ as: 

\begin{equation}
S(h,z)=\alpha(h)S_{\text{max}}(h,z)
\end{equation}
where $S_{\text{max}}\left(h,z\right)~(\text{m}^3 \text{m}^{-3}\text{s}^{-1})$ is the maximum possible root extraction rate when soil water is not limiting and $\alpha(\cdot)$ is a dimensionless water stress reduction factor which is a function of the pressure head  $h~(\text{m})$. $\alpha (h)$ can be expressed as:
\begin{equation}
\alpha(h)=\begin{cases}
0 & \text{$h \leq h_1 $}\\
\frac{h-h_1}{h_2-h_1} &\text{$ h_1 \leq h\leq h_2 $}\\
1& \text{$ h_2\leq h\leq h_3 $}\\
\frac{h_w-h}{h_w-h_3}&\text{$ h_3\leq h\leq h_w $}
\end{cases}
\end{equation}
where $h_1~(\text{m})$ is the pressure head below which roots start to extract water from the soil, $h_2~(\text{m})$ and $h_3~(\text{m})$ are the pressure heads between which optimal water uptake exists, and $h_w~(\text{m})$ is the permanent wilting point pressure head \cite{gupta2016integrative}. In this paper, optimum water uptake is assumed.
The Feddes model consists of two alternative models for $S_{\text{max}}\left(h,z\right)$: independent of depth and depth dependent. The independent of depth model is used in this paper and it is expressed in \cite{Babajimopoulos1995} as:
\begin{equation}
S_{\text{max}}\left(h,z\right)=\frac{\text{TP}_p}{L}
\end{equation}  
where $\text{TP}_p~(\text{ms}^{-1})$ is the potential transpiration rate and $L~(\text{m})$ is the rooting depth.
The potential transpiration rate, $\text{TP}_p~(\text{ms}^{-1})$ is computed by:
\begin{equation}
\text{TP}_p= \text{ETP}_p - \text{EV}
\end{equation} 
where $\text{ETP}_p~(\text{ms}^{-1})$ is potential evapotranspiration (sum of potential transpiration and potential evaporation) and $\text{EV}~(\text{ms}^{-1})$ is the potential evaporation rate computed as in \cite{Al-Khafaf1978} by:
\begin{equation}
\text{EV}=\text{ETP}_p\times \text{exp}(-0.623\text{LAI})
\end{equation}
where $\text{LAI}~(\text{m}^2\text{m}^{-2})$ is the Leaf Area Index function.
Potential evapotranspiration, $\text{ETP}_p~(\text{ms}^{-1})$ can be computed by: 
\begin{equation}
\text{ETP}_p=K_c\text{PET}
\end{equation}
where $\text{PET}~(\text{ms}^{-1})$ is the reference evapotranspiration which is computed from meteorological data using the Penman-Monteith equation and $K_c(-)$ is the crop coefficient. 

\subsection{Cylindrical Coordinates Version of the Richards Equation}
In this paper, to account for the circular movement of the center pivot, we model a field equipped with a center pivot irrigation system with the cylindrical coordinates version of the Richards equation. Specifically, Equation (\ref{eq:Richardseqn_general}) is expressed in terms of $r,~\theta,~\text{and}~z$ where $r~(\text{m})$ represents the radial direction which denotes the radius of the field, $\theta $ represents the azimuthal direction which denotes the angle of rotation of the center pivot, and $z ~ ({\text{m}})$ is the axial direction which represents the depth of soil under consideration. 

In order to express the Richards equation in cylindrical coordinates, the gradient  ($\nabla$) and divergence  ($\nabla.$) operators present in Equation (\ref{eq:Richardseqn_general}) are expressed in their cylindrical forms. 
In cylindrical coordinates, the gradient operator is defined as:
\begin{eqnarray}
\nabla\coloneqq \frac{\partial}{\partial r} \hat{r} ~ + ~ \frac{1}{r}\frac{\partial}{\partial \theta}\hat{\theta}~+~\frac{\partial}{\partial z}\hat{z}
\end{eqnarray}
where $\hat{r}$, $\hat{\theta}$, $\hat{z}$ are the unit vectors in the radial, azimuthal and axial directions respectively.
The divergence of a continuously differentiable vector field $F$ in the cylindrical coordinate system is defined as:
\begin{eqnarray}
\nabla. F \coloneqq \frac{1}{r}\frac{\partial }{\partial r}(rF_r) + \frac{1}{r}\frac{\partial F_\theta}{\partial \theta} +\frac{\partial F_z}{\partial z}
\end{eqnarray}
where $F_r$, $F_\theta$, and $F_z$ are the components of the vector field $F$ in the radial, azimuthal and axial directions respectively.
Applying the operators to the corresponding terms in Equation (\ref{eq:Richardseqn_general}) yields the cylindrical coordinates version of the Richards equation which is expressed as:
\begin{eqnarray}\label{eq:Richards_polar}
C(h)\frac{\partial h}{\partial t}=\frac{1}{r}\frac{\partial}{\partial r}\bigg[r K(h)\frac{\partial h}{\partial r}\bigg]+\frac{1}{r}\frac{\partial}{\partial \theta}\bigg[\frac{K(h)}{r}\frac{\partial h}{\partial \theta}\bigg]+\frac{\partial}{\partial z}\bigg[K(h)\bigg(\frac{\partial h  }{\partial z}+ 1\bigg)\bigg]-S\left(h,z\right)
\end{eqnarray}
Equation (\ref{eq:Richards_polar}) is a nonlinear elliptic-parabolic partial differential equation (PDE) with respect to the spatial ($r,~\theta,~z$) and temporal ($t$) variables respectively.

The soil hydraulic functions $\theta (h)$, $K(h)$ and $C(h)$ in Equation (\ref{eq:Richards_polar}) are described by the Mualem-van Genucthen model \cite{VanGenuchten1980}:
\begin{eqnarray}\label{eq:thetareln}
\theta (h)=\theta_r +(\theta_s-\theta_r)\bigg[\frac{1}{1+(-\alpha h)^n}\bigg]^{1-\frac{1}{n}}
\end{eqnarray}
\begin{equation}
K(h)=K_s\bigg[(1+(-\alpha h)^n)^{-\big(\frac{n-1}{n}\big)}\bigg]^{\frac{1}{2}}\times
\Bigg[1-\bigg[1-\Big[(1+(-\alpha h)^n)^{-\big(\frac{n-1}{n}\big)}\Big]^{\frac{n}{n-1}}\bigg]^{\frac{n-1}{n}}\Bigg]^2
\end{equation}
\begin{equation}
C(h)=(\theta_s-\theta_r)~\alpha n~\bigg[1-\frac{1}{n}\bigg]~(-\alpha h)^{n-1}\big[1+(-\alpha h)^n\big]^{-\big(2-\frac{1}{n}\big)}
\end{equation}
where $\theta_s~(\text{m}^3\text{m}^{-3})$, $\theta_r~(\text{m}^3\text{m}^{-3})$, $K_s~(\text{ms}^{-1})$ are the saturated volumetric moisture content, residual moisture content and saturated hydraulic conductivity, respectively and $n$, $\alpha$ are curve-fitting  soil hydraulic properties.

\subsection{Finite Difference Model Development}
Obtaining an analytical solution to Equation (\ref{eq:Richards_polar}) is difficult due to its nonlinearity, thus we employ a numerical technique to solve the PDE. In this paper, the method of lines (MOL) approach is used. The central difference scheme is used to approximate the derivatives with respect to the spatial variables. The detailed approximation procedure in the $r$, $\theta$, and $z$ directions is described next.
\subsubsection{Discretization in the Radial Direction}

\begin{multline}
\frac{1}{r}\frac{\partial}{\partial r}\bigg[r K(h)\frac{\partial h}{\partial r}\bigg]\bigg|_{e_r,e_\theta,k}\approx  
\frac{1}{r_{e_r,e_\theta,k}\Delta r_i}\bigg[r_{e_r+\frac{1}{2},e_\theta,k}K_{e_r+\frac{1}{2},e_\theta,k}(h)\bigg(\frac{h_{e_r+1,e_\theta,k}-h_{e_r,e_\theta,k}}{\Delta r_E}\bigg)-\\ 
r_{e_r-\frac{1}{2},e_\theta,k}K_{e_r-\frac{1}{2},e_\theta,k}(h)\bigg(\frac{h_{e_r,e_\theta,k}-h_{e_r-1,e_\theta,k}}{\Delta r_W}\bigg)\bigg]
\end{multline}
where $e_r \in [0, N_r+1] $, $e_\theta \in [0, N_\theta+1] $ and $k \in [0, N_z+1] $ represent the position indices in the radial, azimuthal and axial directions respectively. $e_r,~e_\theta,~k ~\subset~\mathbb{Z}_+$. $N_a$ is the total number of nodes (states) in the $a$ direction. $\Delta r_E= r_{e_r+1,e_\theta,k}-r_{e_r,e_\theta,k}$, $\Delta r_W= r_{e_r,e_\theta,k}-r_{e_r-1,e_\theta,k}$ and $\Delta r_i = \frac{1}{2}(\Delta r_E + \Delta r_W)$. $r_{e_r\pm\frac{1}{2},e_\theta,k}=\frac{1}{2}(r_{e_r,e_\theta,k}+r_{e_r\pm1,e_\theta,k})$ and $K_{e_r\pm\frac{1}{2},e_\theta,k}\left(h\right)\approx \frac{1}{2}(K\left(h_{e_r,e_\theta,k}\right)+K\left(h_{e_r\pm 1,e_\theta,k}\right))$. All the nodes in the radial direction are in the center of the $e_r^{\text{th}}$ compartment.

\subsubsection{Discretization in the Azimuthal Direction}

\begin{multline}
\frac{1}{r}\frac{\partial}{\partial \theta}\bigg[\frac{K(h)}{r}\frac{\partial h}{\partial \theta}\bigg]\bigg|_{e_r,e_\theta,k}\approx
\frac{1}{r_{e_r,e_\theta,k}\Delta \theta_j}\bigg[\frac{K_{e_r,e_\theta+\frac{1}{2},k}(h)}{r_{e_r,e_\theta+\frac{1}{2},k}}\bigg(\frac{h_{e_r,e_\theta+1,k}-h_{e_r,e_\theta,k}}{\Delta \theta_T}\bigg)-\\
\frac{K_{e_r,e_\theta-\frac{1}{2},k}(h)}{r_{e_r,e_\theta-\frac{1}{2},k}}\bigg(\frac{h_{e_r,e_\theta,k}-h_{e_r,e_\theta-1,k}}{\Delta \theta_D}\bigg)\bigg]
\end{multline}
where $\Delta \theta_T= \theta_{e_r,e_\theta+1,k}-\theta_{e_r,e_\theta,k}$,      $\Delta \theta_D= \theta_{e_r,e_\theta,k}-\theta_{e_r,e_\theta-1,k}$ and $\Delta \theta_j = \frac{1}{2}(\Delta \theta_T+\Delta \theta_D)$. $r_{e_r,e_\theta \pm\frac{1}{2},k}=\frac{1}{2}(r_{e_r,e_\theta,k}+r_{e_r,e_\theta\pm1,k})=r_{e_r,e_\theta,k}$ and $K_{e_r,e_\theta\pm\frac{1}{2},k}\left(h\right)\approx \frac{1}{2}(K\left(h_{e_r,e_\theta,k}\right)+K\left(h_{e_r,e_\theta\pm 1,k}\right))$.
All the nodes in the azimuthal direction are in the center of the $e_\theta^{\text{th}}$ compartment.

\subsubsection{Discretization in the Axial Direction}
\begin{multline}\label{eq:axialapprox}
\frac{\partial}{\partial z}\bigg[K(h)\bigg(\frac{\partial h  }{\partial z}+ 1\bigg)\bigg]\bigg|_{e_r,e_\theta,k}\approx\frac{1}{\Delta z_k}\bigg[K_{e_r,e_\theta,k+\frac{1}{2}}(h)\bigg(\frac{h_{e_r,e_\theta,k+1}-h_{e_r,e_\theta,k}}{\Delta z_N} +1\bigg)~~~~~\\
-K_{e_r,e_\theta,k-\frac{1}{2}}(h)\bigg(\frac{h_{e_r,e_\theta,k}-h_{e_r,e_\theta,k-1}}{\Delta z_S}+1\bigg)\bigg]
\end{multline}
where $\Delta z_N= z_{e_r,e_\theta,k+1}-z_{e_r,e_\theta,k}$,  $\Delta z_S= z_{e_r,e_\theta,k}-z_{e_r,e_\theta,k-1}$, $\Delta z_k = \frac{1}{2}(\Delta z_N+\Delta z_S)$ and  $K_{e_r,e_\theta,k\pm\frac{1}{2}}\left(h\right)\approx\frac{1}{2}(K\left(h_{e_r,e_\theta,k}\right)+K\left(h_{e_r,e_\theta,k\pm 1}\right))$. All the nodes in the axial direction are in the center of the $k^{\text{th}}$ compartment.
The resulting ODE, in terms of the temporal variable, is obtained expressed as:
\begin{multline}\label{eq:finitedifference}
\frac{dh}{dt}=\frac{1}{C_{e_r,e_\theta,k}(h)}\Bigg[\Bigg(\frac{1}{r_{e_r,e_\theta,k}\Delta r_i}\bigg[r_{e_r+\frac{1}{2},e_\theta,k}K_{e_r+\frac{1}{2},e_\theta,k}(h)\bigg(\frac{h_{e_r+1,e_\theta,k}-h_{e_r,e_\theta,k}}{\Delta r_E}\bigg)- \\~~~~~~~~~~~~~~~
r_{e_r-\frac{1}{2},e_\theta,k}K_{e_r-\frac{1}{2},e_\theta,k}(h)\bigg(\frac{h_{e_r,e_\theta,k}-h_{e_r-1,e_\theta,k}}{\Delta r_W}\bigg)\bigg]\Bigg)+\\
~~~~~~~~~\Bigg(\frac{1}{r_{e_r,e_\theta,k} \Delta \theta_j}\bigg[\frac{K_{e_r,e_\theta+\frac{1}{2},k}(h)}{r_{e_r,e_\theta+\frac{1}{2},k}}\bigg(\frac{h_{e_r,e_\theta+1,k}-h_{e_r,e_\theta,k}}{\Delta \theta_T}\bigg)-\\
\frac{K_{e_r,e_\theta-\frac{1}{2},k}(h)}{r_{e_r,e_\theta-\frac{1}{2},k}}\bigg(\frac{h_{e_r,e_\theta,k}-h_{e_r,e_\theta-1,k}}{\Delta \theta_D}\bigg)\bigg]\Bigg)+\\
~~~~~~~\Bigg(\frac{1}{\Delta z_k}\bigg[K_{e_r,e_\theta,k+\frac{1}{2}}(h)\bigg(\frac{h_{e_r,e_\theta,k+1}-h_{e_r,e_\theta,k}}{\Delta z_N} +1\bigg)~~~~~\\
-K_{e_r,e_\theta,k-\frac{1}{2}}(h)\bigg(\frac{h_{e_r,e_\theta,k}-h_{e_r,e_\theta,k-1}}{\Delta z_S}+1\bigg)\bigg]\Bigg)- S\left(h_{e_r,e_\theta,k},z\right)\Bigg]
\end{multline}

\subsubsection{Treatment of the Axis}
In this paper, L'Hopital's rule is applied to the first and second terms at the right-hand side of Equation (\ref{eq:Richards_polar}) to treat the singularity that occurs at the axis ($r=0$). The resulting equations are expressed as follows:
\vspace{-2mm}
\begin{equation}\label{eq:sing_r}
\lim_{r\to 0} \frac{1}{r}\frac{\partial}{\partial r}\bigg[r K(h)\frac{\partial h}{\partial r}\bigg] = \lim_{r\to 0}\frac{\frac{\partial }{\partial r}\left(\frac{\partial}{\partial r}\bigg[r K(h)\frac{\partial h}{\partial r}\bigg]\right)}{\frac{\partial }{\partial r}\left( r \right)} = 2\frac{\partial }{\partial r}\left(K\left(h\right)\frac{\partial h}{\partial r}\right)  
\end{equation}

\begin{equation}\label{eq:sing_t}
\lim_{r\to 0} \frac{1}{r}\frac{\partial}{\partial \theta}\bigg[\frac{K(h)}{r}\frac{\partial h}{\partial \theta}\bigg]=\lim_{r\to 0}\frac{\frac{\partial}{\partial r}\left(\frac{\partial}{\partial \theta}\bigg[K(h)\frac{\partial h}{\partial \theta}\bigg]\right)}{\frac{\partial }{\partial r}\left(r^2\right)} =\frac{1}{2}\frac{\partial }{\partial r} \left( \frac{\partial }{\partial r}\left(\frac{\partial}{\partial \theta}\bigg[K(h)\frac{\partial h}{\partial \theta}\bigg]\right)\right)
\end{equation}
Substitution of Equations (\ref{eq:sing_r}) and (\ref{eq:sing_t}) into Equation (\ref{eq:Richards_polar}) yields the cylindrical coordinates version of the Richards equation at $r=0$:
\vspace{-2mm}

\begin{equation}\label{eq:finalsingularity}
C(h)\frac{\partial h}{\partial t}=2\frac{\partial }{\partial r}\left(K\left(h\right)\frac{\partial h}{\partial r}\right)+\frac{1}{2}\frac{\partial }{\partial r} \left( \frac{\partial }{\partial r}\left(\frac{\partial}{\partial \theta}\bigg[K(h)\frac{\partial h}{\partial \theta}\bigg]\right)\right)+\frac{\partial}{\partial z}\bigg[K(h)\bigg(\frac{\partial h  }{\partial z}+ 1\bigg)\bigg]-S\left(h,z\right)
\end{equation}
The central difference approximations of the spatial derivatives in Equations (\ref{eq:sing_r}) and (\ref{eq:sing_t}) are expressed as follows:
\vspace{-2mm}
\begin{equation}
2\frac{\partial }{\partial r}\left(K\left(h\right)\frac{\partial h}{\partial r}\right) \approx 2\frac{\left(K_{e_r + \frac{1}{2}, e_\theta, k}\left(h\right)\left[h_{e_r + 1, e_\theta, k} -  h_{e_r, e_\theta, k}\right] - K_{e_r -\frac{1}{2}, e_\theta, k}\left(h\right)\left[h_{e_r, e_\theta, k} -  h_{e_r-1, e_\theta, k}\right]\right)}{\Delta r_i ^2}
\end{equation}
\vspace{-6mm}
\begin{multline}
\frac{1}{2}\frac{\partial }{\partial r} \left( \frac{\partial }{\partial r}\left(\frac{\partial}{\partial \theta}\bigg[K(h)\frac{\partial h}{\partial \theta}\bigg]\right)\right)\approx \frac{1}{2\Delta r_i^2 \Delta \theta_k^2}\bigg[ \bigg(K_{e_r+1, e_\theta + \frac{1}{2}, k}\left(h\right)\left[h_{e_r + 1, e_\theta+1, k} -  h_{e_r+1, e_\theta, k}\right]-\\
~~~~~~~~~~~~~~~~~~~~~~~~~~~~~~~~~~~~~~~~~~~~~~~~~~~~~K_{e_r+1, e_\theta-\frac{1}{2}, k}\left(h\right)\left[h_{e_r + 1, e_\theta, k} -  h_{e_r+1, e_\theta-1, k}\right]\bigg)-\\ 
~~~~~~~~~~~~~~~~~~~~~~~~~~~~~~~~~~~~~~~~~~~~~~~~~~~2\bigg(K_{e_r, e_\theta + \frac{1}{2}, k}\left(h\right)\left[h_{e_r, e_\theta+1, k} -  h_{e_r, e_\theta, k}\right]-\\
 ~~~~~~~~~~~~~~~~~~~~~~~~~~~~~~~~~~~~~~~~~~~~~~~~~~K_{e_r, e_\theta-\frac{1}{2}, k}\left(h\right)\left[h_{e_r, e_\theta, k} -  h_{e_r, e_\theta-1, k}\right]\bigg)+
 \\ ~~~~~~~~~~~~~~~~~~~~~~~~~~~~~~~~~~~~~~~~~~~~~~~~\bigg(K_{e_r-1, e_\theta + \frac{1}{2}, k}\left(h\right)\left[h_{e_r-1, e_\theta+1, k} -  h_{e_r-1, e_\theta, k}\right] -\\
  K_{e_r-1, e_\theta - \frac{1}{2}, k}\left(h\right)\left[h_{e_r-1, e_\theta, k} -  h_{e_r-1, e_\theta-1, k}\right]\bigg)\bigg]
\end{multline}

\subsubsection{Initial and Boundary Conditions}
Equation (\ref{eq:Richards_polar}) is solved numerically for the following initial and boundary equations which apply to fields equipped with a center pivot irrigation system:
\begin{align}
h(r,\theta,z, t=0)&= h_{\text{int}}\\
\frac{\partial h\left(r,\theta, z, t\right)}{\partial r}\bigg|_{\left(r=0,~\theta,~z\right)}&=0 \\
\frac{\partial h\left(r,\theta, z, t\right)}{\partial r}\bigg|_{\left(r=H_r,~\theta,~z\right)}&=0 \label{eq:leftBC}\\
\frac{\partial h\left(r,\theta, z, t\right)}{\partial \theta}\bigg|_{\left(r=0,~\theta,~z\right)}&=0 \\
\frac{\partial (h\left(r,\theta,z,t\right)+z)}{\partial z}\bigg|_{\left(r,~\theta,~z=0\right)}&=1 \\
\frac{\partial (h\left(r,\theta,z,t\right))}{\partial z}\bigg|_{\left(r,~\theta,~z=H_z\right)}&=-1-\frac{u_{\text{irr}}}{K(h)}\label{eq:topBC}\\
h(r,~\theta =0,~z,~t)&=h(r,~\theta=2\pi,~z,~t)\label{eq:revBC}
\end{align}
where $H_r$ in Equation (\ref{eq:leftBC}) is the total radius of the field, $H_z$ in Equation (\ref{eq:topBC}) is the length of the soil column and $u_{\text{irr}}$ (m/s) in Equation (\ref{eq:topBC}) represents the irrigation rate which is considered as the input in this paper.

\subsection{State-space Representation of the Field Model}
The field model is expressed in state space form as:
\begin{eqnarray}\label{eq:statespace}
\dot{x}=f(x(t),u(t)) + \omega(t)
\end{eqnarray}
where $x(t)\in \mathbb{R}^{N_x}$ represents the state vector containing $N_x$ pressure head values for the corresponding spatial nodes, defined at time instant $t$. $u(t)\in \mathbb{R}^{N_u}$ and $\omega(t)\in \mathbb{R}^{N_x}$ represent the input and the model disturbances respectively. The general output function, taking into account measurement noise is expressed as:
\begin{eqnarray}\label{eq:output}
y(t)=h(x(t),t) + v(t)
\end{eqnarray}
where $y(t)\in \mathbb{R}^{N_y}$, $v(t)\in \mathbb{R}^{N_y}$ respectively denote the measurement vector and the measurement noise. Equation (\ref{eq:output}) is the general form of Equation (\ref{eq:thetareln}).

\section{State Estimator Design}
In this paper, the discrete-time extended Kalman filter (EKF) is chosen as the state estimator. The EKF is a well-known method used for the estimation of nonlinear systems based on a successive linearization of the nonlinear system about the previous estimate of the states. The EKF is made up of two main steps, the prediction, and the update steps. The prediction step updates the state $x$ and its covariance matrix $P$ using the model of the system. The update step is carried out when measurements are available. In the presence of measurements, $x$ and $P$ are updated again by first computing the gain matrix $K$. The detailed steps are shown below.

\textit{Initialization}. The continuous-time system, Equations (\ref{eq:statespace}) and (\ref{eq:output}), is discretized to obtain its discrete-time equivalent. The discrete-time version can be expressed as:
\begin{eqnarray*}
x_{k+1}=F(x(k),u(k)) + \omega(k)\\
y_k=H(x(k),k) + v(k)
\end{eqnarray*}
and the filter is initialized as follows:
\begin{gather*}
E[x_0]=\hat{x}_0 \\
E[(x_0-\hat{x}_0)(x_0-\hat{x}_0)^T]=P(0|0)
\end{gather*}

\textit{Prediction Step}. Given the previous state estimate $\hat{x}_{k|k}$, the previous observation sequence $Y_k$ and the new input $u_k$ to the system, the new state of the system at time $t_{k+1}$:
\begin{gather*}
\hat{x}_{k+1|k}=E[x_{k+1}|Y_k,u_k]\\~~~~
=F(\hat{x}_{k|k},u_k)
\end{gather*}
where $Y_k\coloneqq \{y_0, y_1, ....., y_k\}$. The state covariance matrix can be calculated as:
\begin{gather*}
~~~~~~~~~~~~~~~~~~P(k+1|k)=E[(x_{k+1}-\hat{x}_{k+1|k})(x_{k+1}-\hat{x}_{k+1|k})^T|Y_k,u_k]\\=A_kP(k|k)A_k^T+Q
\end{gather*}
where $A_k=\frac{\partial F}{\partial x}\big|_{\hat{x}_{k|k},u_k}$ and $Q$ is the covariance matrix of the process disturbance $\omega$.

\textit{Filtering Step}. In this step, we use the observation $y_{k+1}$ at time $t_{k+1}$ to update the state and its covariance. The Kalman gain matrix, $K_{k+1}$ calculation
\begin{equation*}
K_{k+1}=P(k+1|k)H^T_{k+1}[H_{k+1}P(k+1|k)H^T_{k+1} + R]^{-1}
\end{equation*}
where $H_{k+1}=\frac{\partial H}{\partial x}\big|_{\hat{x}_{k+1|k}}$ and $R$ is the covariance matrix of the measurement noise $v$. After obtaining the updated Kalman gain, the state is updated:
\begin{gather*}
\hat{x}_{k+1|k+1}=E[x_{k+1}|Y_{k+1},u_k]\\~~~~~~~~~~~~~~~~~~~~~~~~~~~~~~~~~
=\hat{x}_{k+1|k} + K_{k+1}[y_{k+1}-H(\hat{x}_{k+1|k})]
\end{gather*}
where $Y_{k+1}\coloneqq\{Y_k,y_{k+1}\}$. The state covariance update is updated accordingly:
\begin{gather*}
~~~~~~~~~P(k+1|k+1)=E[(x_{k+1}-\hat{x}_{k+1|k+1})(x_{k+1}-\hat{x}_{k+1|k+1})^T|Y_{k+1}]\\=[I-K_{k+1}H_{k+1}]P\left(k+1|k\right)
\end{gather*}

\section{Simulated Case Study}
In this section, the performance of the proposed information fusion system is demonstrated with simulated microwave sensor measurements under the following scenarios: (i) Scenario 1: uniform initial conditions and uniform soil parameters; and (ii) Scenario 2: uniform initial conditions and two layers of soil in the investigated field. We first describe the system from which the simulations are based on. A description of how the simulated microwave sensor measurements are generated is also provided. We then provide results to demonstrate the effectiveness of the proposed information system under the earlier mentioned scenarios.

\subsection{System Description}
\begin{figure}[H]
	\centering
	\centerline{\includegraphics[trim={0cm 0cm 0cm 0cm}, width=0.5\textwidth]{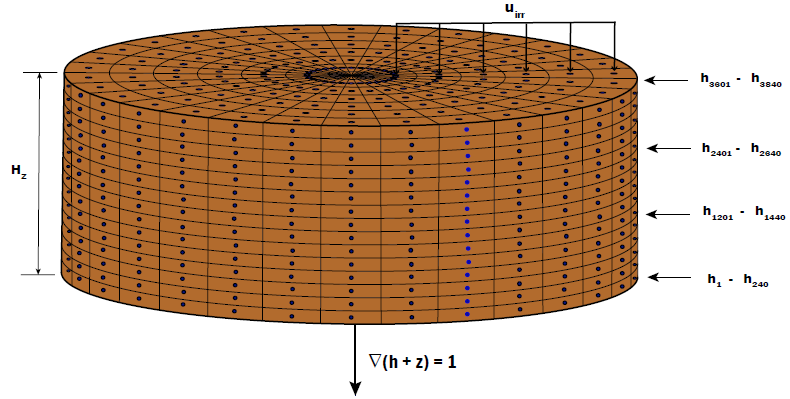}}
	\caption{A schematic diagram of the investigated field.}
	\label{fig:investigated_field}
\end{figure}

\begin{figure}[!ht]
	\centering
	\centerline{\includegraphics[trim={0cm 0cm 0cm 0cm}, width=0.4\textwidth]{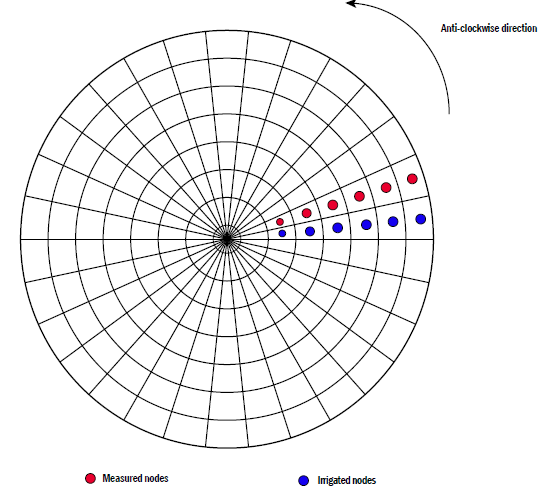}}
	\caption{A schematic representation of center pivots equipped with microwave sensors.}
	\label{fig:simulated_measurements}
\end{figure}

A field of radius 50 m and depth of 0.30 m is investigated in this case study. The field is divided into 6, 40 and 16 compartments in the radial, azimuthal and axial directions respectively. A schematic diagram of the investigated field is shown in Figure \ref{fig:investigated_field}.
In both scenarios, the center pivot moves at a speed of 0.022 m/s. An irrigation amount of 7 mm/day is applied between 0:00 AM to 4:00 AM daily. The soil moisture measurements that are fed into the EKF at each sampling time are obtained from the adjacent sector (represented with the red dots in Figure \ref{fig:simulated_measurements}), in the anti-clockwise direction, of the presently irrigated sector (represented with the blue dots in Figure \ref{fig:simulated_measurements}). Since the presently irrigated sector changes after a specific time, the measured nodes also change. This arrangement describes the operation of microwave sensors mounted on center pivots. 11 soil moisture measurements are used in the update step of the EKF at each sampling time.

This simulated case study considers the presence of a healthy barley crop at its development stage in the investigated field. Thus, crop coefficient values between 0.75 and 0.96 are used in this paper. Daily reference evapotranspiration values of 1.2 mm/day, 1.70 mm/day, 0.6 mm/day, 0.5 mm/day and 2.10 mm/day are used.
Process noise and measurement noise are considered in the simulations and they have zero mean and standard deviations of $1\times 10^{-5}$ m and $1\times 10^{-4}~\text{m}^3 \text{m}^{-3}$ respectively. 
Trajectories of the actual process states and the state estimates calculated by EKF are also compared for some selected nodes. The selected nodes are located at depths of 0.00 m (surface of the field), 0.06 m, 0.12 m, 0.18 m, 0.24 m, and 0.30 m (bottom of the investigated field). The actual process states and the state estimates are converted into volumetric moisture content using Equation (\ref{eq:thetareln}). The absolute error ($e_k$) between the actual volumetric moisture content and the estimated volumetric moisture content is also computed.
\begin{equation}
e_k=|y_k-\hat{y}_k|
\end{equation} 
Moisture content maps are constructed to represent the calculated volumetric moisture content and absolute error. We refer to moisture content maps that are constructed with the actual volumetric moisture content, the estimated volumetric soil moisture content, and the absolute error as the actual, estimated, and absolute error moisture content maps respectively.
\subsection{Results}
\subsubsection{Scenario 1}
In this scenario, a single soil type and uniform initial conditions are considered in the investigated field. Specifically, it is assumed that the investigated field contains loamy soil. Table \ref{tbl:true_value}  shows the initial condition and the hydraulic properties of loamy soil used in the simulations. The EKF is initialized with $\hat{x}_0 = 1.2\times x_0$.

\begin{table}[H]
	\caption{The initial condition and parameters of loamy soil.}
	\small 
	\centering
	\begin{tabular}{cccccc}
		\hline
		 {$x_0$ (m)}& {$K_{s}$ (m/s)} &{$\theta_{s}$ $(\text{m}^{3}/\text{m}^{3})$}&{$\theta_{r}$ $(\text{m}^{3}/\text{m}^{3})$}& {$\alpha$ (1/m)}& {$n$ (-)}\\
		\hline
		 $-$0.8 & $2.889\times 10^{-6}$  & 0.430 & 0.0780&3.60&1.56\\
		\hline
	\end{tabular} \label{tbl:true_value}
\end{table}

\begin{figure}[H]
	\centerline{\includegraphics[trim={0cm 0cm 0cm 0cm}, width=0.6\textwidth]{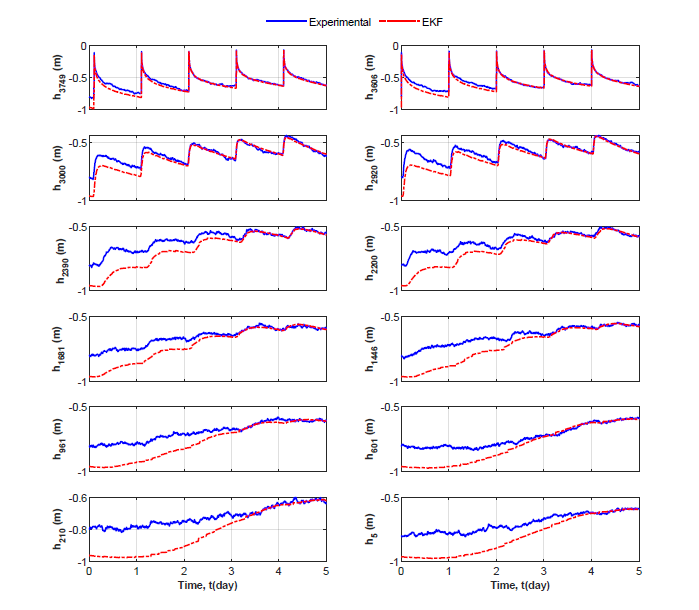}}
	\caption{Some selected trajectories of the actual process states (blue solid line) and the state estimates (red dash-dot) in Scenario 1.}
	\label{fig:SameInit_OneParNorm}
\end{figure}
From Figure \ref{fig:SameInit_OneParNorm}, it can be seen that the EKF (red dash-dot) is able to track the actual process states (blue solid line) very well. Thus, the EKF is able to provide accurate state estimates.
Next, the moisture content maps  constructed at selected times during the simulation period, for the surface and the bottom of the field, are examined.

\begin{figure}[H]
\centering
\includegraphics[width=1\textwidth]{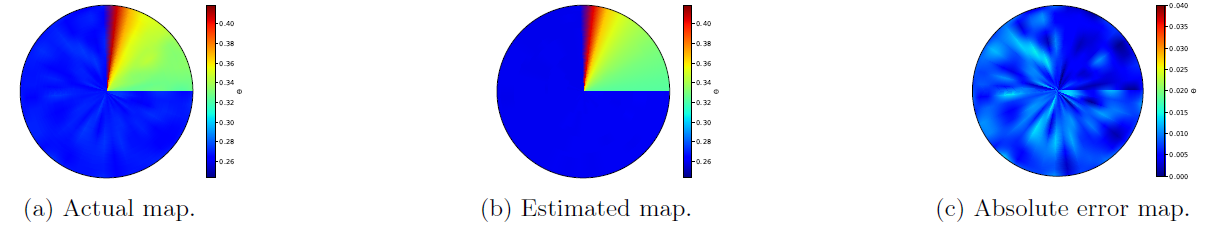}
\caption{Surface moisture content maps at 01:00 HRS on Day 2 in Scenario 1. }
\label{fig:surface_1}
\end{figure}

\begin{figure}[H]
\centering
\includegraphics[width=1\textwidth]{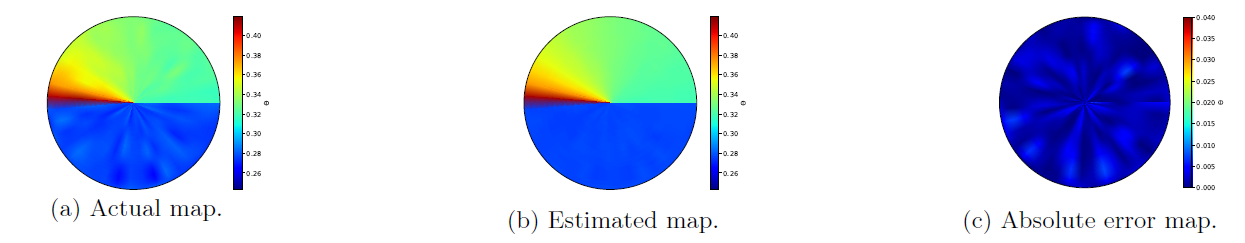}
\caption{Surface moisture content maps at  02:00 HRS on Day 4 in Scenario 1.}
\label{fig:surface_2}
\end{figure}

\begin{figure}[H]
\centering
\includegraphics[width=1\textwidth]{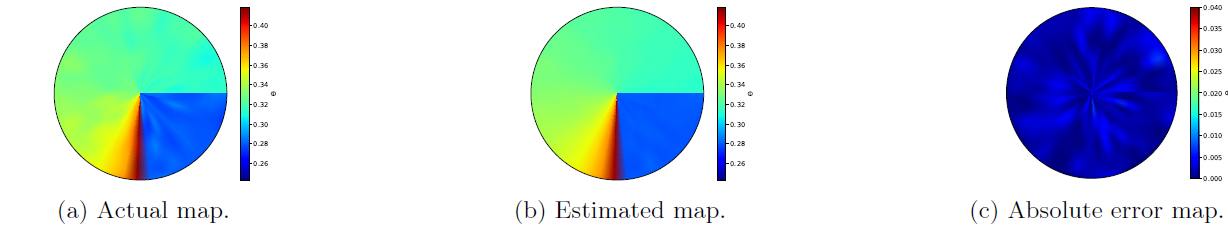}
\caption{Surface moisture content maps at  03:00 HRS on Day 5 in Scenario 1.}
\label{fig:surface_3}
\end{figure}

\begin{figure}[H]
\centering
\includegraphics[width=1\textwidth]{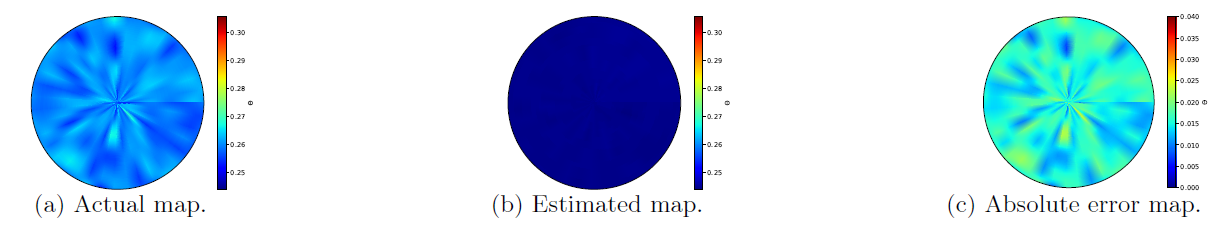}
\caption{Bottom moisture content maps on Day 1 at 01:00 HRS in Scenario 1.}
\label{fig:bottom_1}
\end{figure}

\begin{figure}[H]
\centering
\includegraphics[width=1\textwidth]{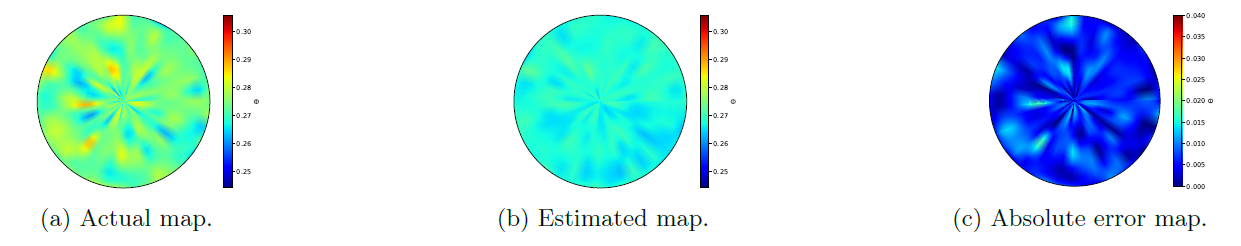}
\caption{Bottom moisture content maps on Day 4 at 02:00 HRS in Scenario 1.}
\label{fig:bottom_2}
\end{figure}

\begin{figure}[H]
\centering
\includegraphics[width=1\textwidth]{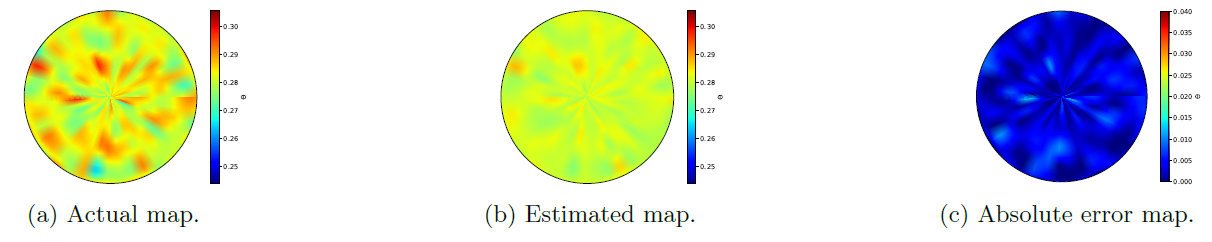}
\caption{Bottom moisture content maps on Day 5 at 03:00 HRS in Scenario 1.}
\label{fig:bottom_3}
\end{figure}

From the surface (Figures \ref{fig:surface_1}, \ref{fig:surface_2}, and \ref{fig:surface_3} ) and the bottom maps (Figures \ref{fig:bottom_1}, \ref{fig:bottom_2}, and \ref{fig:bottom_3}), it is evident that there is a strong agreement between the estimated maps and the actual maps as the simulation proceeds. Thus, the information fusion system is able to provide accurate moisture content maps when the field is made up of uniform soil parameters. 

\subsubsection{Scenario 2}
In nature, soils are seldom uniform over significant length scales, and layered soils are ubiquitous~\cite{farthing2017numerical}. Thus, in this scenario, a field made up of two soil layers is considered. Specifically, the investigated field is made of a loamy layer and a sandy clay loam layer. The loamy layer occupies a depth of 0.16 m while the sandy clay loam layer occupies a depth of 0.14 m and the loamy layer is above the sandy clay loam layer.
The hydraulic properties of sandy clay loam are shown in Table \ref{tbl:true_value2}. 
\begin{table}[H]
	\caption{The hydraulic parameters of sandy clay loam soil.}
	\small 
	\centering
	\begin{tabular}{cccccc}
		\hline
		{$x_0$ (m)}& {$K_{s}$ (m/s)} &{$\theta_{s}$ $(\text{m}^{3}/\text{m}^{3})$}&{$\theta_{r}$ $(\text{m}^{3}/\text{m}^{3})$}& {$\alpha$ (1/m)}& {$n$ (-)}\\
		\hline
		$-$0.8	    & $7.222\times 10^{-7}$  & 0.410 & 0.090&1.90&1.31\\

		\hline
	\end{tabular} \label{tbl:true_value2}
\end{table}

In layered soils, the volumetric water content is discontinuous across the interface between the layers because of the unique capillary head relationships in different soils of layers~\cite{assouline2013infiltration}. Rather, the pressure head is continuous across the layers~\cite{farthing2017numerical}.
Thus in this scenario, it is suitable to directly use the pressure head values at the measured nodes as the output, $y$, in the update step of the EKF. 11 pressure head measurements from the loamy layer are used in the update step of the EKF at each sampling time. The EKF  is initialized with $\hat{x}_0 = 1.2\times x_0$.

\begin{figure}[H]
	\centerline{\includegraphics[trim={0cm 0cm 0cm 0cm}, width=0.60\textwidth]{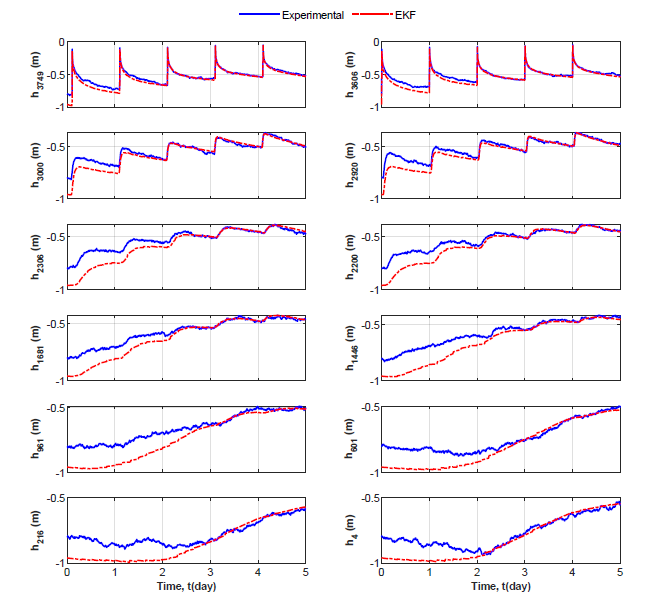}}
	\caption{Some selected trajectories of the actual process states (blue solid line) and the state estimates (red dash-dot) in Scenario 2.}
	\label{fig:TwoPars_Normal}
\end{figure}

From Figure \ref{fig:TwoPars_Normal}, it is evident that the EKF provides accurate estimates when a layered field is considered. As shown in Figure \ref{fig:TwoPars_Normal}, the estimates (red dash-dot) of the states that are located in the sandy clay loam layer ($\text{h}_4$, $\text{h}_{216}$, $\text{h}_{601}$, $\text{h}_{961}$, $\text{h}_{1446}$, $\text{h}_{1681}$) are able to track their corresponding true process states (blue solid line) very well. A similar observation is made for the estimates of the states that are located in the loam layer ($\text{h}_{2200}$, $\text{h}_{2306}$, $\text{h}_{2920}$, $\text{h}_{3000}$, $\text{h}_{3606}$, $\text{h}_{3749}$). The moisture content maps are analyzed in the sequel.

\begin{figure}[H]
\centering
\includegraphics[width=1\textwidth]{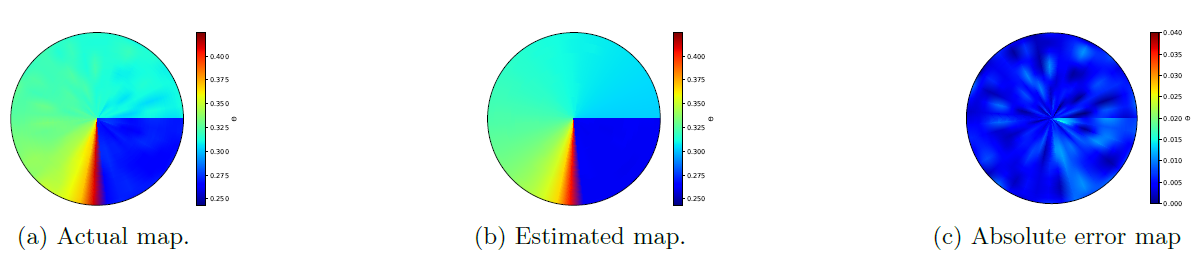}
\caption{Surface moisture content maps (loamy layer) at 03:00 HRS on Day 2 in Scenario 2.}
\label{fig:loam_1}
\end{figure}

\begin{figure}[H]
\centering
\includegraphics[width=1\textwidth]{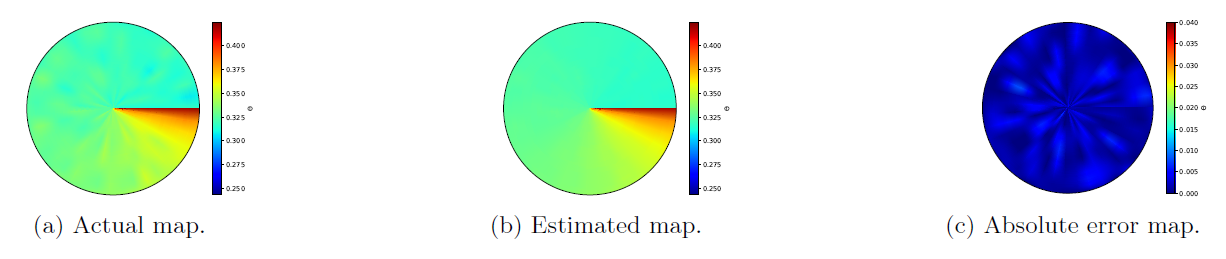}
\caption{Surface moisture content maps (loamy layer) at 04:00 HRS on Day 4 in Scenario 2.}
\label{fig:loam_2}
\end{figure}

\begin{figure}[H]
\centering
\includegraphics[width=1\textwidth]{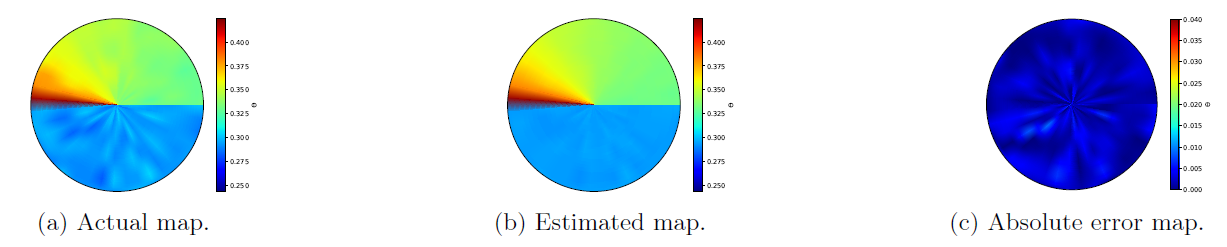}
\caption{Surface moisture content maps (loamy layer) at 02:00 HRS on Day 5 in Scenario 2.}
\label{fig:loam_3}
\end{figure}

\begin{figure}[H]
\centering
\includegraphics[width=1\textwidth]{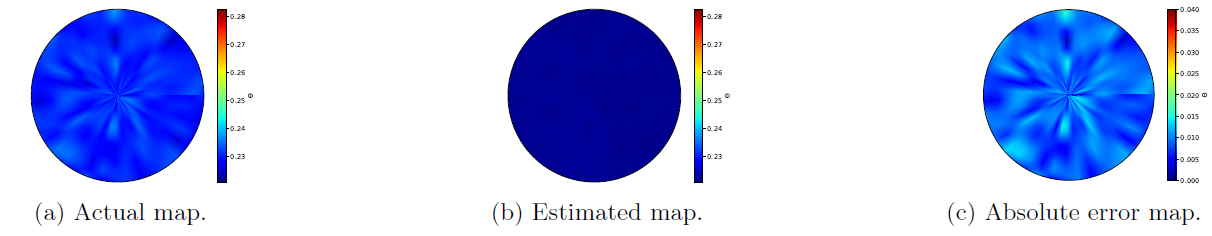}
\caption{Bottom moisture content maps (sandy clay loam layer) on Day 2 at 3:00 HRS in Scenario 2.}
\label{fig:scl_1}
\end{figure}

\begin{figure}[H]
\centering
\includegraphics[width=1\textwidth]{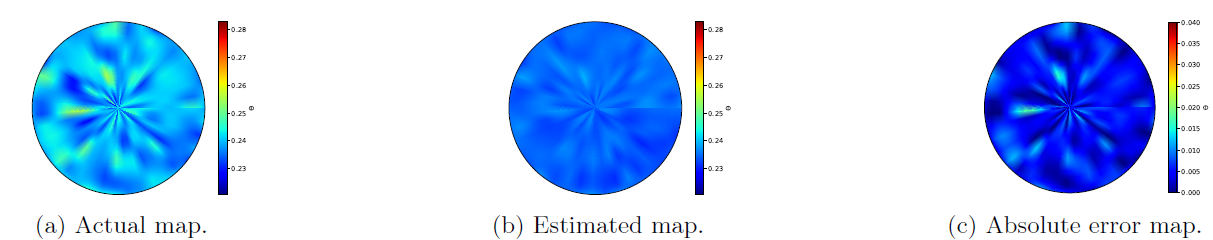}
\caption{Bottom moisture content maps (sandy clay loam layer) on Day 4 at 04:00 HRS in Scenario 2.}
\label{fig:scl_2}
\end{figure}

\begin{figure}[H]
\centering
\includegraphics[width=1\textwidth]{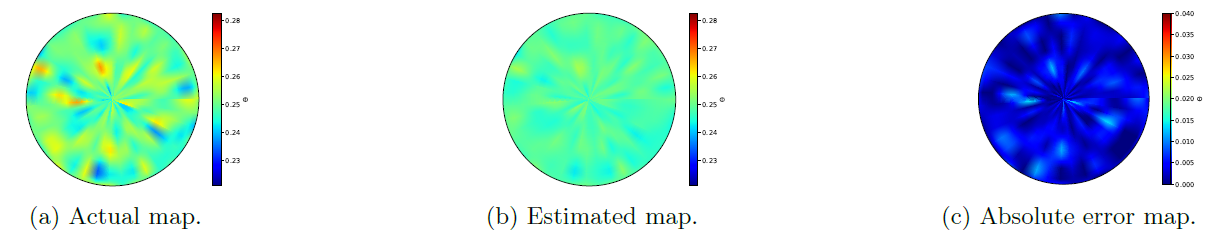}
\caption{Bottom  moisture content maps (sandy clay loam layer) on  Day 5 at 02:00 HRS in Scenario 2.}
\label{fig:scl_3}
\end{figure}
Similar to the uniform soil parameters scenario, the information  fusion system is able to provide moisture content maps that agree with the actual moisture content maps. This is evident in the moisture content maps constructed for the loamy layer (Figures \ref{fig:loam_1}, \ref{fig:loam_2}, and \ref{fig:loam_3}) and the sandy clay loam layer (Figures \ref{fig:scl_1}, \ref{fig:scl_2}, and \ref{fig:scl_3}). The information fusion system is thus able to provide accurate moisture content maps when the investigated field is made up of different soil types.
\section{Real Case Study}
 In this section, we demonstrate the utility, and performance of the proposed information fusion system by considering microwave remote sensor measurements obtained from a field equipped with a center pivot irrigation system. Essentially, the modeling and estimation methods described in Sections 2 and 3 are used in this section. Firstly, a description of the study area is provided. Secondly, a series of data preprocessing steps, which result in a suitable data representation for the EKF are briefly discussed. Thirdly, we introduce the criteria for evaluating the performance of the EKF and hence the information fusion system. Lastly, we present and discuss the results of the numerical investigation.

\subsection{Field Description}
\begin{figure}[H]
	\centerline{\includegraphics[width=0.55\textwidth]{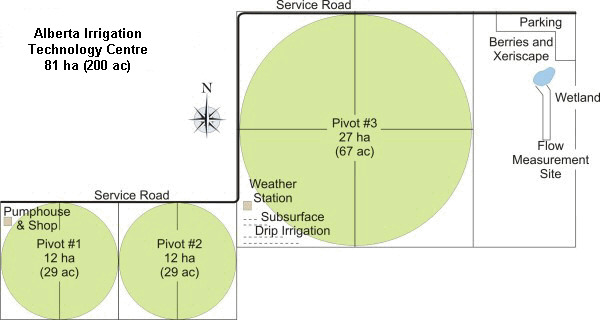}}
	\caption{Layout of the Alberta Irrigation Center.} \vspace{2mm}
	\label{fig:layout_of_the_alberta_irrigation_center_}
\end{figure}

The Alberta Irrigation Center is located east of the City of Lethbridge, Alberta, Canada, with an approximate area of 0.81 $\text{km}^2$. The main soil texture is clayey loam with few lenses of sand within the soil profile. The layout of the center is shown in Figure \ref{fig:layout_of_the_alberta_irrigation_center_}.
\vspace{2mm}
The field under study is the one covered by Pivot $\#3$. The Lethbridge Demo Farm Irrigation Management Climate Information Network (IMCIN) provides meteorological parameters such as temperature, incoming solar radiation, and wind. Real-time data from this weather station can be obtained from the Alberta Climate Information Service website (https://agriculture.alberta.ca/acis/).

\subsection{Data Preprocessing}
The measurements used in this study go through a series of data preprocessing steps to obtain a suitable data representation for the EKF. The various preprocessing steps are enumerated and explained below.

\begin{enumerate}[label={(\arabic*)}]
\item \textbf{Discarding measurements beyond the circular track of the center pivot}: It is considered that only measurements on and within the circular track of the center pivot are useful. Thus, all measurements whose locations fall beyond the circular track are identified and excluded.

\item \textbf{Sorting measurements by date and time}: After excluding the measurements that fall beyond the circular track of the pivot, the retained data is sorted by date. Measurements obtained on a specific day are further sorted by time in ascending order.

\item \textbf{Sorting measurements by quadrants}: To make the field model computationally tractable, especially for large fields, we propose the division of the field model into four submodels, with each submodel representing one of the four quadrants of the entire field. Consequently, the measurements obtained on a particular day are sorted by the quadrants in which they are located.

\item \textbf{Center pivot movement detection}: The measurements in each quadrant are grouped according to a specific sampling time such that the change in measurement locations with time models the anticlockwise movement of the center pivot.

\item \textbf{Outlier detection}: For a given soil type present in a  quadrant, moisture measurements that are lower than the residual moisture content, $\theta_r$, are discarded. Similarly, moisture measurements greater than the saturated moisture content, $\theta_s$, are also discarded.

\item \textbf{Mapping measurements to nodes in the field model}: To map each preprocessed measurement to a node in the field model, we first generate GPS coordinates whose arrangement is similar to the arrangement of nodes in the field model. For an incoming measurement, the distances between its GPS coordinates and the generated coordinates are computed and the measurement is assigned to the node which has the least distance from the incoming measurement location.
\end{enumerate}

\subsection{Evaluation Criteria}
To evaluate the performance of the information fusion system in providing consistent and accurate soil moisture estimates and predictions, the following evaluation methods are employed:

\begin{enumerate}[label={(\arabic*)}]
	\item \textbf{Cross-validation}: The measurements obtained at each sampling time are divided into two sets: a training set and a validation set. The training set is used in the update step of the EKF. Measurements in the validation set are compared with their corresponding estimates provided by the EKF.

	\item \textbf{The Normalized Innovation Squared (NIS) test}: In practice, the performance of the EKF can be measured by comparing the predicted and true values of the measurements. The measurement residual, known as the \textbf{innovation} $e_k$, is defined as the difference between the actual measurement, $y_k$, and the best available prediction based on the system model and previous measurements, $\hat{y}_k$. Mathematically, $e_k$ is defined as:
\begin{gather}
e_k=y_k-\hat{y}_k\\~~~
\hat{y}_k=H(\hat{x}_{k|k-1})
\end{gather}
where $H(\cdot)$ is the output function and $\hat{x}_{k|k-1}$ is \textit{a priori}.
The covariance matrix, $E_k$ of $e_k$ is: 
\begin{eqnarray}
E_k=H_kP(k|k-1)H_k^T + R
\end{eqnarray}
where $R$ is the measurement noise covariance matrix, $P(k|k-1)$ is the covariance matrix of the state $\hat{x}_{k|k-1}$ and $H_k=\frac{\partial H}{\partial x}\big|_{\hat{x}_{k|k-1}}$.
The EKF is working properly if $e_k$ is Gaussian, zero-mean, white (uncorrelated), and with covariance equal to $E_k$. To verify this, a statistical test term is defined in terms of $E_k$ and $e_k$. This statistic is known as the normalized innovation squared (NIS) \cite{bar2004estimation} and it is defined as:
\begin{eqnarray}
\Omega^2_{d,k}=e_k^T E_k^{-1} e_k
\end{eqnarray}
A null hypothesis which supposes that the sample observations result purely from chance can be formulated as: 
\begin{eqnarray}
H_0: E(e_k)=0
\end{eqnarray}
$H_0$ can be tested against the alternative hypothesis:
\begin{eqnarray}
H_1: E(e_k)\neq 0
\end{eqnarray}
$\Omega^2_{d,k}$ follows a $\chi ^2$ distribution with the probability relationship:
\begin{eqnarray}
P\{\Omega^2_{d,k}\leq\chi^2_{m,1-\alpha}|H_0\}=1-\alpha
\end{eqnarray}
where the subscript $m$, the degrees of freedom, corresponds to the number of observations at time instant $k$ and $\alpha$ is the significance level.

If $\Omega^2_{d,k}<\chi^2_{m,1-\alpha}$, the null hypothesis can be accepted \cite{gamse2014statistical}. We can conclude that there is no significant discrepancy between the system estimate and the measurement model.
	\item \textbf{Evolution of the state covariance matrix's trace}: In the EKF, the presence of measurements reduces the uncertainty of the state estimates. The trace of the state covariance matrix ($P$) quantifies the uncertainty of the state estimates. As a means of testing the performance of the EKF, the trace of the state covariance matrix is monitored during the numerical investigation.
\end{enumerate}

\subsection{System Description}
To adapt the Equation (\ref{eq:finitedifference}) to a sector of the circular track, two boundary conditions are imposed in the $\theta$-direction, and the boundary condition that is expressed by Equation (\ref{eq:revBC}) is omitted. Specifically, the zero-gradient boundary condition is imposed at the boundaries of the sector in the $\theta$-direction.
\begin{eqnarray}
\frac{1}{r}\frac{\partial h(r,\theta,z,t)}{\partial \theta}=0~~~~~~~ at~~~(r,\theta=\theta_1,z) 
\end{eqnarray}
\begin{eqnarray}
\frac{1}{r}\frac{\partial h(r,\theta,z,t)}{\partial \theta}=0~~~~~~~ at~~~(r,\theta=\theta_2,z) 
\end{eqnarray}

\begin{figure}[H]
	\centerline{\includegraphics[trim={0cm 0cm 0cm 0cm}, width=0.6\textwidth]{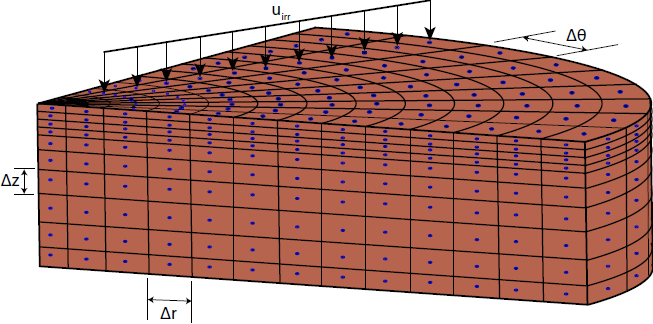}}
	\caption{A schematic diagram of the investigated quadrant.}
	\label{fig:investigated_quadrant}
\end{figure}

One of the quadrants is chosen for the numerical investigation. We point out that although the proposed approach targets the entire field, only one quadrant is chosen for the sake of simplicity and with no loss of generality. Indeed, the current approach can be applied to the remaining quadrants by considering the field conditions and the microwave measurements obtained in those quadrants.
The investigated quadrant has a radius ($H_r$) of $290$ m and the center pivot describes a total angle ($H_{\theta}$) of $0.5\pi$ radians to fully traverse it. The depth of soil ($H_z$) investigated is $0.6$ m. $H_r$ and $H_{\theta}$ are equally divided into $30$ and $17$ equally spaced compartments respectively. $H_z$ is divided into $10$ unequally spaced compartments with a finer discretization near the soil surface and a coarsening discretization away from the soil surface. Correspondingly, the field submodel representing Quadrant 4 is discretized into a total of $5,100$ states. A schematic diagram of the investigate quadrant is shown in Figure \ref{fig:investigated_quadrant}.

\begin{figure}[H]
\centering
\includegraphics[width=1\textwidth]{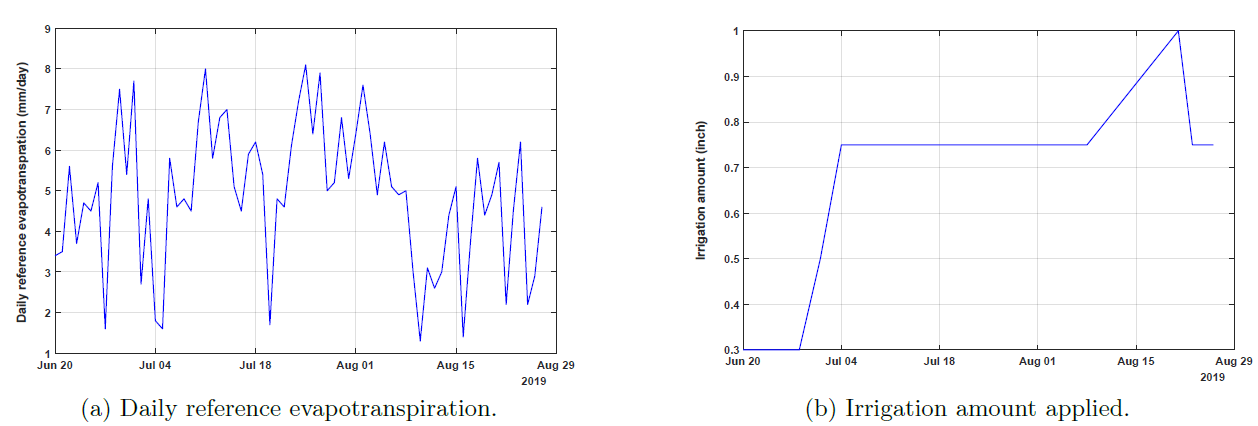}
\caption{Weather conditions and applied irrigation from June $20^{\text{th}}$ - August $27^{\text{th}}$, 2019.}
\label{fig:observed_weather_conditions_and_applied_irrigation}
\end{figure}

Measurements provided by the microwave sensors from June $20^{\text{th}}$, 2019 to August $27^{\text{th}}$, 2019 are considered. Crop coefficient values of barley between 0.6 and 1.5 are incorporated into the investigation. The irrigation applied during the period is shown in Figure \ref{fig:observed_weather_conditions_and_applied_irrigation}(b). The soil hydraulic parameters of clay loam, the soil type in the investigated quadrant, and the initial guess of the state are shown in Table 4.2. Weather information in the form of reference evapotranspiration for the period under investigation was obtained from the ACIS website and the observed values are shown in Figure \ref{fig:observed_weather_conditions_and_applied_irrigation}(a).

\subsection{Results}

\subsubsection{Frequently Updated Moisture Content Maps}
\begin{figure}[H]
\includegraphics[width=1\textwidth]{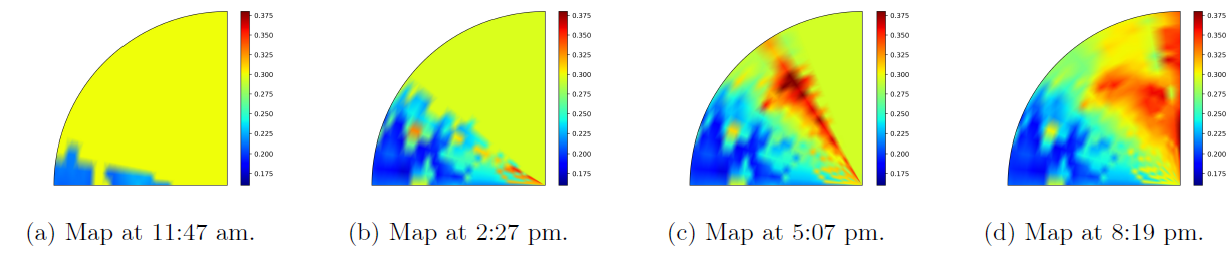}
\caption{Frequently updated moisture content maps from
	11:47 am to 8:19 pm
		on June $20^{\text{th}}$.}
\label{fig:frequently_constructed}
\end{figure}

We employ the near-surface maps obtained on June $20^{\text{th}}$ from 11:47 am to 8:19 pm to illustrate the construction of the frequently updated maps. The measurements  are sampled every 32 minutes. At each sampling time, the microwave measurements are assimilated into the field model using the EKF in the proposed information fusion system. The state estimates provided by the EKF are converted into volumetric moisture using Equation (\ref{eq:thetareln}). The resulting volumetric moisture estimates are represented with moisture content maps.
When new measurements are obtained, the soil moisture estimates and hence the moisture content maps are updated. This process is repeated until all the measurements in the selected quadrant and on a particular day have been considered. The process of frequently updating the volumetric moisture estimates and hence the moisture content maps is illustrated with Figures \ref{fig:frequently_constructed}(a) to \ref{fig:frequently_constructed}(d).

\subsubsection{Moisture Content Maps for Selected Days}
\begin{figure}[H]
\centering
\includegraphics[width=1\textwidth]{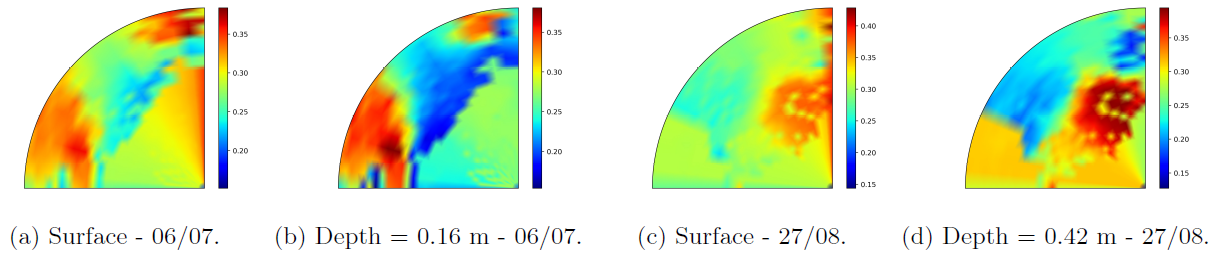}
\caption{Maps at the surface and greater depths below the surface.}
\label{fig:other_days}
\end{figure}

We show moisture content maps generated on some selected days (July $6^{\text{th}}$ at 21:55 pm, and August $27^{\text{th}}$ at 15:31 pm) within the simulation period. In addition to the near-surface maps, maps at greater depths below the soil surface are also shown.
Figure \ref{fig:other_days}(a) and Figure \ref{fig:other_days}(c) respectively depict the near-surface maps after all the measurements on July $6^{\text{th}}$ and August $27^{\text{th}}$ were assimilated into the field model. On $6^{\text{th}}$ July, the moisture map at the depth of 0.16 m from the soil surface is shown in Figure \ref{fig:other_days}(b). The map at a depth of 0.42 m from the soil surface on the $27^{\text{th}}$ of August is shown in Figure \ref{fig:other_days}(d).

\subsubsection{Predictive Capability of the Proposed Approach}
\begin{figure}[H]
\centering
\includegraphics[width=1\textwidth]{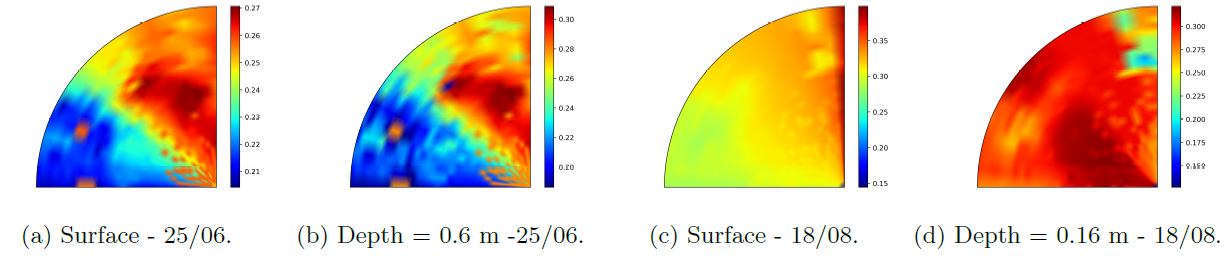}
\caption{Moisture content maps constructed with soil moisture predictions.}
\label{fig:moisture_content_maps_predictions}
\end{figure}

Figure \ref{fig:moisture_content_maps_predictions} shows moisture content maps constructed on days within the simulation period when the center pivot was stationary. These maps are constructed with soil moisture predictions provided by the field model. Figure \ref{fig:moisture_content_maps_predictions}(a) shows the predicted near-surface soil water content on $25^{\text{th}}$ June and Figure \ref{fig:moisture_content_maps_predictions}(b) shows the predicted soil water content at a depth of 0.60 m.
Similarly, on the $18^{\text{th}}$ of August, the predicted near-surface soil moisture content is shown in Figure \ref{fig:moisture_content_maps_predictions}(c). Figure \ref{fig:moisture_content_maps_predictions}(d) provides a summary of the predicted soil water content at a depth of 0.16 m.

\subsubsection{Trace of the State Covariance Matrix}
\begin{figure}[H]
\centering
\includegraphics[trim={0 0cm 0cm 0cm},width=0.6\textwidth]{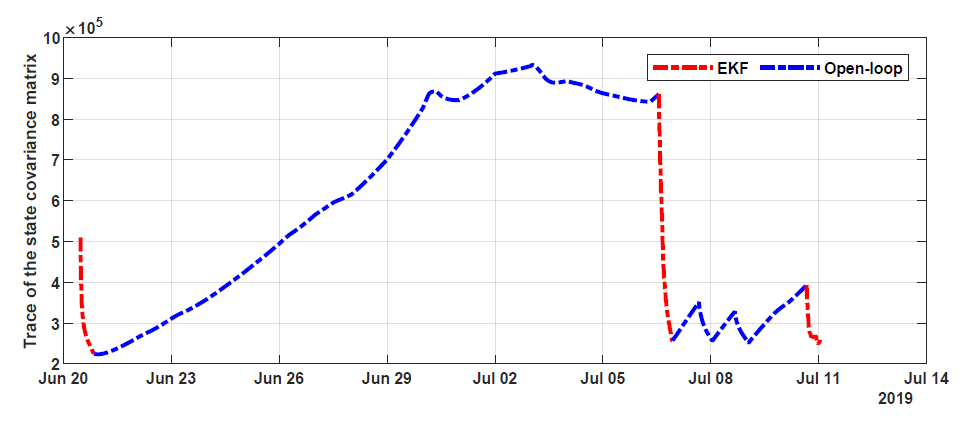}
\caption{Trajectory of the trace of the state covariance matrix from June $20^{\text{th}}$ to July $11^{\text{th}}$.}
	\label{fig:trajectory_of_the_trace}
\end{figure}

Figure \ref{fig:trajectory_of_the_trace} shows the trajectory of the trace of the state covariance matrix from June $20^{\text{th}}$ to July $11^{\text{th}}$. From Figure \ref{fig:trajectory_of_the_trace}, it is evident that in the presence of microwave measurements (EKF update), there is a decrease (red dash-dot segments) in the trace of the state covariance matrix. Thus, when measurements were available from the microwave sensors, there is a continuous reduction in the uncertainty of the soil moisture estimates.

\subsubsection{Results of the NIS test}
\begin{figure}[H]
\centering
\includegraphics[trim={0 0cm 0cm 0cm},width=0.9\textwidth]{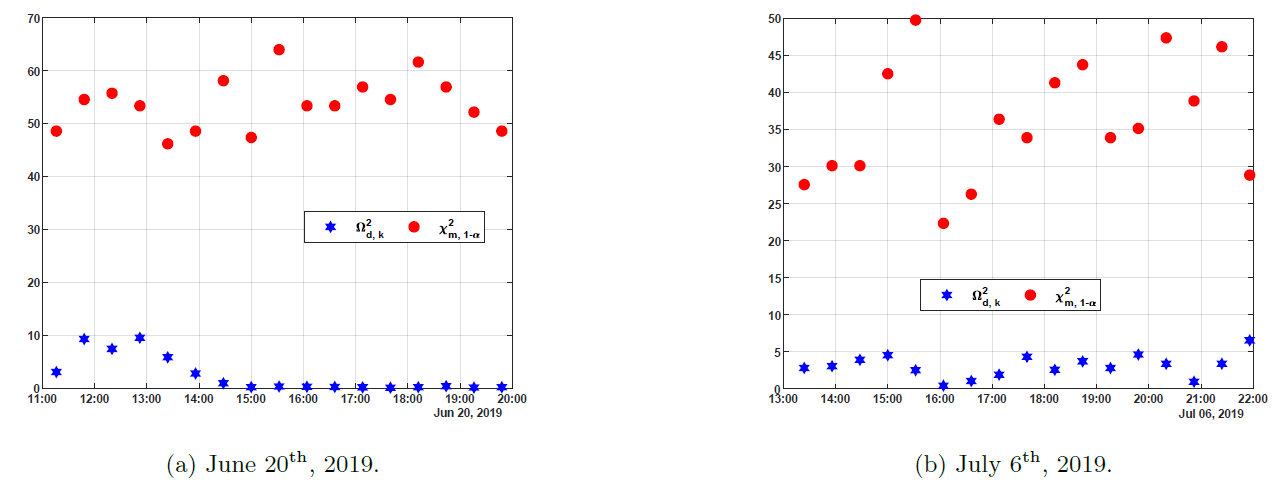}
\caption{A comparison of the NIS test statistic (blue hexagon) and $\chi^2_{m,1-\alpha}$ (red dots) for two selected days within the period under investigation.}
	\label{fig:nis_test_results}
\end{figure}

Figure \ref{fig:nis_test_results} shows the results of the NIS test for some selected days (June $20^{\text{th}}$ and July $6^{\text{th}}$) performed at a 5\% significance level. From Figure \ref{fig:nis_test_results}, it is evident that, at each time instant $t_k$, $\Omega^2_{d,k} ~\text{(blue hexagon)}$ is less than $\chi^2_{m,1-0.05} ~\text{(red dots)}$. We can therefore accept the null hypothesis and conclude that there is no significant discrepancy between the state estimates provided by the proposed information fusion system and the measurements obtained from the microwave sensors.

\subsubsection{Results of the Cross-validation}
\begin{figure}[H]
\centering
\includegraphics[trim={0 0cm 0cm 0cm},width=1\textwidth]{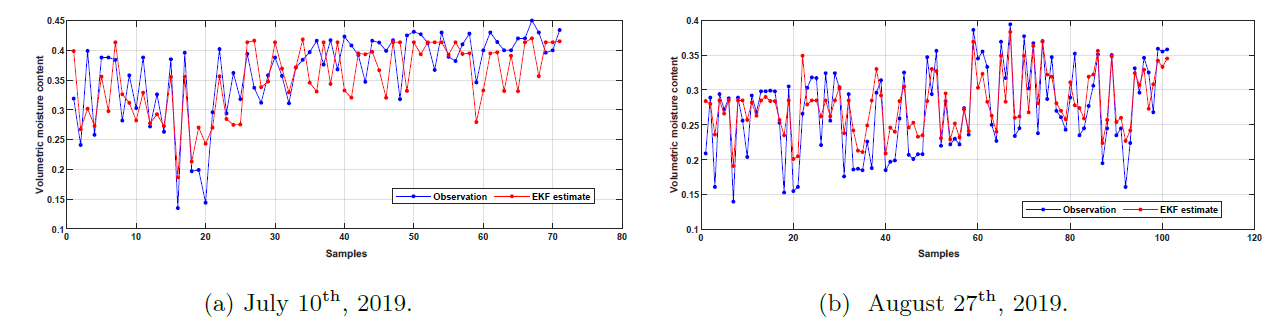}
\caption{ A comparison of the measurements in the validation set (blue dots) with their corresponding estimates (red dots) for two  selected days with the period under investigation.}
\label{fig:_cross_validation_type1}
\end{figure}

\begin{figure}[H]
\centering
\includegraphics[trim={0 0cm 0cm 0cm},width=0.8\textwidth]{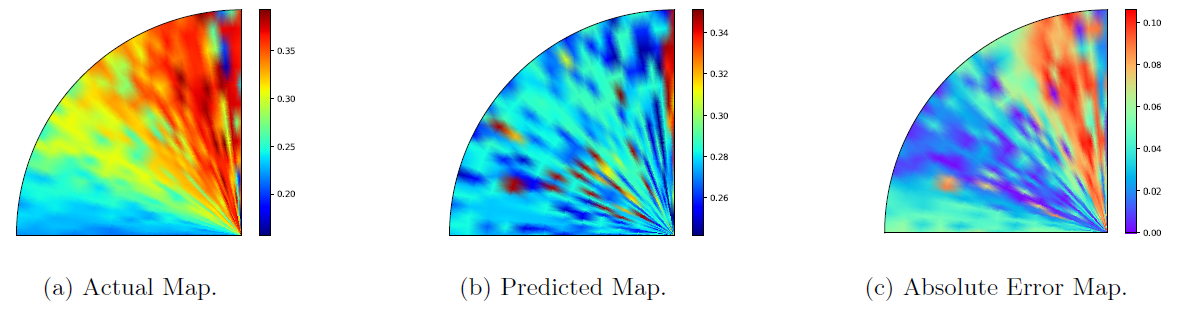}
\caption{A comparison between the soil moisture predictions and the true measurements obtained on August  $13^{\text{th}}$, 2019.}
	\label{fig:validation_results_type2}
\end{figure}

Figures \ref{fig:_cross_validation_type1} and \ref{fig:validation_results_type2} show the results of the cross-validation. Two kinds of cross-validation were performed. 
In the first kind, shown in Figure \ref{fig:_cross_validation_type1}, the cross-validation is performed on a fraction (validation set) of the total preprocessed measurements obtained on the day under consideration. In this approach, we randomly split the total measurements into a training set and a validation set, in a ratio of 80\% to 20\%. From Figures \ref{fig:_cross_validation_type1}(a) and \ref{fig:_cross_validation_type1}(b), it is observed that the approximate average absolute error between the actual observations and their corresponding estimates is 0.049 and 0.0273 for July $10^{\text{th}}$ and August $27^{\text{th}}$ respectively. The soil moisture estimates capture the trend in the measurements present in the validation set. Consequently, it can be concluded that the EKF estimates are within negligible error bounds from the true measurements. 

In the second kind, the cross-validation is performed with all the preprocessed measurements obtained on the $13^{\text{th}}$ of August. This was done by simulating the field model, taking into consideration the applied irrigation and the prevailing weather conditions observed on the $13^{\text{th}}$ of August. The soil moisture predictions resulting from the simulation, shown in Figure \ref{fig:validation_results_type2}(b) are then compared with the actual measurements, shown in Figure \ref{fig:validation_results_type2}(a). From the absolute error map, Figure \ref{fig:validation_results_type2}(c), it is evident that close to 80\% of the absolute error values are less than 0.06 and the maximum absolute error obtained is 0.10. Thus, the field model can provide soil moisture predictions that are within practical error bounds from the actual measurements obtained from the microwave sensors. From the results of the cross-validation, we can conclude that the soil moisture predictions and estimates provided by the information fusion system are reliable and accurate.

\section{Conclusions}
In this paper, the challenges associated with soil moisture sensing using microwave remote sensors mounted on center pivots are addressed and a moisture map construction procedure that is suitable for the implementation of closed-loop irrigation is proposed. We propose the cylindrical coordinates version of the Richards equation to model fields equipped with center pivots. The microwave sensor measurements are then assimilated into the resulting model using the EKF to form an information fusion system. Water content maps are constructed with the soil moisture estimates and predictions obtained from the proposed information fusion system. The results obtained in the simulated case study confirmed that accurate soil moisture estimates and hence accurate water content maps can be obtained from the EKF and the proposed information fusion system. The real case study demonstrated the ability of our approach to effectively deal with the challenges. Frequently updated maps are constructed to eliminate the time delays associated with the current microwave sensing approach and the field model provided soil moisture predictions when the center pivot is stationary. By integrating the microwave sensor measurements with the field model, soil moisture information at greater depths below the surface can be inferred. The results of the performance evaluation confirmed that the soil moisture estimates and predictions obtained from the information fusion system are consistent and accurate. 

\section*{Acknowledgment}
Financial support from Alberta Innovates and Natural Sciences and Engineering Research Council of Canada is gratefully acknowledged.

\end{document}